\documentclass[aps,preprint,showpacs,preprintnumbers,eqsecnum,amsmath,amssymb]{revtex4}

\usepackage{graphics,epsfig}
\usepackage{graphicx}
\usepackage{dcolumn}
\usepackage{bm}

\begin{document}

\title{Dirac quasinormal frequencies of Reissner-Nordstr\"om black hole
in Anti-de Sitter spacetime}
\author{Jiliang Jing  and Qiyuan Pan} \email{jljing@hunnu.edu.cn}
\affiliation{ Institute of Physics and  Department of Physics, \\
Hunan Normal University,\\ Changsha, Hunan 410081, P. R. China }

 \baselineskip=0.65 cm

\vspace*{0.2cm}
\begin{abstract}
\vspace*{0.2cm}

The quasinormal modes (QNMs) of Dirac field perturbations of a
Reissner-Nordstr\"om black hole in an asymptotically Anti-de
Sitter spacetime are investigated. We find that both the real and
imaginary parts of the fundamental quasinormal frequencies for
large black holes are the linear functions of the Hawking
temperature, and the slope of the lines for the real parts
decreases while that for the magnitude of the imaginary parts
increases as the black hole charge increases. According to the
Anti-de Sitter/Conformal Field Theory (AdS/CFT) correspondence,
the fact shows that different charge presents different time scale
in three-dimensional CFT. Another interesting result is that the
quasinormal frequencies become evenly spaced for high overtone
number, and in the spacing expressions the real part decreases
while the magnitude of the imaginary part increases as the charge
increases. We also study the relation between quasinormal
frequencies and angular quantum number and find that the real part
increases while the magnitude of the imaginary part decreases as
the angular quantum number increases.

\end{abstract}

 \vspace*{1.5cm}

\maketitle

\section{introduction}

It is well known that QNMs possess discrete spectra of complex
characteristic frequencies which are entirely fixed by the
structure of the background spacetime and are irrelevant of the
initial perturbations\cite{Chand75} \cite{Frolov98}. Thus, it is
believed that one can directly identify a black hole existence by
comparing QNMs with the gravitational waves observed in the
universe, as well as test the stability of the event horizon
against small perturbations. Meanwhile, it is generally believed
that the study of QNMs may lead to a deeper understanding of the
thermodynamic properties of black holes in loop quantum gravity
\cite{Hod} \cite{Dreyer} since the real part of quasinormal
frequencies with a large imaginary part for the scalar field in
the Schwarzschild black hole is equal to the Barbero-Immirzi
parameter \cite{Hod}\cite{Dreyer} \cite{Baez}\cite{Kunstatter}, a
factor introduced by hand in order that the loop quantum gravity
reproduces correct entropy of the black hole.

Motivated by the AdS/CFT correspondence, the study of QNMs in AdS
spacetime becomes appealing recently. It was argued that the QNMs
of AdS black holes have a direct interpretation in terms of the
dual conformal field theory \cite{Maldacena} \cite{Witten}
\cite{Kalyana}. According to the AdS/CFT correspondence, a large
static black hole in asymptotically AdS spacetime corresponds to
an (approximately) thermal state in CFT, and the decay of the test
field in the black hole spacetime corresponds to the decay of the
perturbed state in CFT. The dynamical timescale for the return to
the thermal equilibrium can be done in AdS spacetime, and then
translated onto the CFT, using the AdS/CFT correspondence. The
QNMs in AdS spacetime were first computed for a conformally
invariant scalar field by Chan and Mann \cite{Chan}. Subsequently,
Horowitz and Hubeny \cite{Horowitz} presented a numerical method
to evaluate the quasinormal frequencies directly and made a
systematic investigation of QNMs for the scalar field in the
Schwarzschild AdS black hole. Then, many authors studied QNMs for
the scalar, electromagnetic and gravitational perturbations in
various asymptotically AdS black holes \cite{Cardoso1}
-\cite{Kurita}. Although QNMs for the scalar, electromagnetic and
gravitational field perturbations in the Schwarzschild AdS and the
Reissner-Nordstr\"om AdS black-hole backgrounds have been
investigated extensively, QNMs for the Dirac field perturbations
in the Schwarzschild AdS black hole were first studied in Refs.
\cite{Max}\cite{Jing} recently. Considering the
Reissner-Nordstr\"om AdS (RNAdS) black hole spacetime provides a
better background than the Schwarzschild AdS geometry, in this
paper we will extend the study in Ref. \cite {Jing} to the RNAdS
black hole and see how the black hole charge affects the Dirac
QNMs.

The organization of this paper is as follows. In Sec.2 the
decoupled Dirac equations and the corresponding wave equations in
the RNAdS spacetime are obtained by using Newman-Penrose
formalism. In Sec.3 the numerical approach to compute the Dirac
QNMs is introduced. In Sec.4 the numerical results for the Dirac
QNMs in the RNAdS black hole are presented. The last section is
devoted to a summary.

\section{Dirac equations in the Reissner-Nordstr\"om Anti-de Sitter spacetime}

The line element of the RNAdS black hole can be expressed as
\begin{eqnarray} \label{metric}
ds^2=f dt^2-\frac{1}{f}dr^2-r^2(d\theta^2+sin^2\theta d\varphi^2),
\end{eqnarray}
with
\begin{eqnarray}
f=1-\frac{2 M}{r}+\frac{Q^2}{r^2}+\frac{r^2}{R^2},
\end{eqnarray}
where $M$, $Q$ and $R$ represent the mass, charge and anti-de
Sitter radius, respectively. Hereafter we will take $R=1$ and
measure everything in terms of $R$. The spacetime causal structure
depends on the zeros of $f$. Changing the parameters $M$ and $Q$,
the function $f$ may have none, one or two positive zeros.  In
general case, $f$ has two simple real, positive roots $r_-$ and
$r_+$, but in the so-called extreme case, $f$ has one double
positive zero $r_+$ (because $r_-=r_+$). The horizons $r_-$ and
$r_+$ with $r_-<r_+$, are called Cauchy and event horizons
respectively. The Hawking temperature of the black hole is given
by
\begin{eqnarray}
T_H=\frac{\kappa}{2\pi}=\frac{1-Q^2/r_+^2+3 r_+^2}{4\pi r_+ },
\end{eqnarray}
where $\kappa$ is the surface gravity. The mass parameter $M$ is
related to the charge $Q$ and the event horizon radius $r_+$ by
the relation
\begin{eqnarray}
M=\frac{1}{2}\left(r_++r_+^3+\frac{Q^2}{r_+}\right),
\end{eqnarray}
and the extremal value of the black hole charge, $Q_{ext}$, is
given by
\begin{eqnarray}
Q^2_{ext}=r_+^2 \left(1+3 r_+^2\right).
\end{eqnarray}

In a curved spacetime the Dirac equations \cite{Page} are
described by
\begin{eqnarray}
   &&\sqrt{2}\nabla_{BB'}P^B+i\mu \bar{Q}_{B'}=0, \nonumber \\
   &&\sqrt{2}\nabla_{BB'}Q^B+i\mu \bar{P}_{B'}=0,
\end{eqnarray}
where $\nabla_{BB'}$ is covariant differentiation, $P^B$ and $Q^B$
are the two-component spinors representing the wave function, $
\bar{P}_{B'}$ is the complex conjugate of $P_{B}$, and $\mu $ is
the particle mass. In order to separate the Dirac equations in the
RNAdS spacetime (\ref{metric}) by using Newman-Penrose formulism
\cite{Newman}, we take the null tetrad as
\begin{eqnarray}
  &&l^\mu=(\frac{r^2}{\Delta}, ~1, ~0, ~0 ), \nonumber \\
  &&n^\mu=\frac{1}{2}(1, ~-\frac{\Delta}{r^2}, ~0, ~0)\nonumber \\
  &&m^\mu=\frac{1}{\sqrt{2} r}\left(0, ~0, ~1, \frac{i}{sin\theta}\right),
\end{eqnarray}
with
\begin{eqnarray}
\Delta=r^2-2M r+Q^2+r^4,
\end{eqnarray}
and set the two-component spinors as
\begin{eqnarray}
&&P^0=\frac{1}{r}{\mathbb{R}}_{-1/2}S_{-1/2}e^{-i(\omega t-m\varphi)}, \nonumber \\
&&P^1={\mathbb{R}}_{+1/2}S_{+1/2}e^{-i(\omega t-m\varphi)}, \nonumber \\
&&\bar{Q}^{1'}={\mathbb{R}}_{+1/2}S_{-1/2}e^{-i(\omega t-m\varphi)}, \nonumber \\
&&\bar{Q}^{0'}=-\frac{1}{r}{\mathbb{R}}_{-1/2}S_{+1/2}e^{-i(\omega
t-m\varphi)},
\end{eqnarray}
where ${\mathbb{R}}_{+1/2}$ and ${\mathbb{R}}_{-1/2}$ are
functions of the coordinate $r$, and $S_{+1/2}$ and  $S_{-1/2}$
are functions of the coordinate $\theta$.  Following the steps in
Refs. \cite{Jing, Jing2}, after the tedious calculation we find
that the Dirac equations can be reduced to the following radial
and angular parts
\begin{eqnarray}\label{dd2}
&&\sqrt{\Delta}{\mathcal{D}}_0 {\mathbb{R}}_{-1/2}=(\lambda+i\mu
r)\sqrt{\Delta} {\mathbb{R}}_{+1/2}, \\
\label{dd3}&&\sqrt{\Delta}{\mathcal{D}}_0^{\dag}
(\sqrt{\Delta}{\mathbb{R}}_{+1/2})=(\lambda-i\mu r) {\mathbb{R}}_{-1/2},\\
&&{\mathcal{L}}_{1/2} S_{+1/2}=-\lambda S_{-1/2}, \label{aa1}\\
&&{\mathcal{L}}_{1/2}^{\dag} S_{-1/2}=\lambda S_{+1/2},\label{aa2}
\end{eqnarray}
with
 \begin{eqnarray}
 &&{\mathcal{D}}_n=\frac{\partial}{\partial r}-\frac{i K}
 {\bigtriangleup}+\frac{n}{\bigtriangleup}\frac{d \Delta}{d r},\nonumber \\
 &&{\mathcal{D}}^{\dag}_n=\frac{\partial}{\partial r}+\frac{i K}
 {\bigtriangleup}+\frac{n}{\bigtriangleup}\frac{d \Delta}{d r},\nonumber \\
 &&{\mathcal{L}}_n=\frac{\partial}{\partial \theta}+\frac{m}{\sin \theta }
 +n\cot \theta,\nonumber \\
 &&{\mathcal{L}}^{\dag}_n=\frac{\partial}{\partial \theta}-\frac{m}{\sin \theta }
 +n\cot \theta, \nonumber \\
 &&K=r^2\omega.\label{ld}
 \end{eqnarray}
We can eliminate $S_{+1/2}$ (or $S_{-1/2}$) from Eqs. (\ref{aa1})
and (\ref{aa2}) and obtain
\begin{eqnarray}\label{ang}
\left[\frac{1}{sin\theta}\frac{d}{d
\theta}\left(sin\theta\frac{d}{d\theta}\right)-\frac{m^2+2mscos\theta+s^2cos^2\theta}
{sin^2\theta}+s+A_s\right]S_s=0,
\end{eqnarray}
where $A_{+1/2}=\lambda^2-1$ and $A_{-1/2}=\lambda^2$.  The
angular equation (\ref{ang}) can be solved exactly and
$A_s=(l-s)(l+s+1)$, where $l$ is the quantum number characterizing
the angular distribution. Thus, for both cases $s=\pm 1/2$ we have
 \begin{eqnarray}
\lambda^2=\left(l+\frac{1}{2}\right)^2.
 \end{eqnarray}

In what follows, we focus our attention on the massless Dirac
field. Then, we can eliminate ${\mathbb{R}}_{-1/2}$ (or
${\mathbb{R}}_{+1/2}$) from Eqs. (\ref{dd2}) and (\ref{dd3}) to
obtain a radial decoupled Dirac equation for $
{\mathbb{R}}_{+1/2}$ (or ${\mathbb{R}}_{-1/2}$), and we find that
both them can be expressed as
\begin{eqnarray}  \label{T1}
&& \Delta^{-s} \frac{d }{d r}\left(\Delta^{1+s}
\frac{d{\mathbb{R}}_{s}}{d r}\right)+P{\mathbb{R}}_{s}
 =0,
 \end{eqnarray}
with
 \begin{eqnarray}
&&P=\frac{K^2-is K\frac{d \Delta}{dr}}{ \Delta} +4s i  \omega
r+\frac{1}{2}\left(s+\frac{1}{2}\right)\frac{d^2 \Delta}{d r^2}
-\lambda^2.
 \end{eqnarray}
Introducing an usual tortoise coordinate
 \begin{eqnarray}
dr_*=(r^2/\Delta) dr,
 \end{eqnarray}
 and resolving the equations in the form
 \begin{eqnarray}
&&{\mathbb{R}}_{s}=\frac{\Delta^{-s/2}}{r} \Psi_s,
 \end{eqnarray}
we obtain
\begin{eqnarray}\label{wave}
\frac{d^2 \Psi_s }{d r_*^2}+(\omega ^2-V )\Psi_s =0,
\end{eqnarray}
where
\begin{eqnarray}\label{Poten}
V=-\frac{\Delta}{4 r^2}\frac{d}{d
r}\left[r^2\frac{d}{dr}\left(\frac{\Delta}{r^4}\right)\right]+\frac{s^2
r^4}{4}\left[\frac{d}{d
r}\left(\frac{\Delta}{r^4}\right)\right]^2+is \omega r^2\frac{d}{d
r}\left(\frac{\Delta}{r^4}\right)+\frac{\lambda^2 \Delta}{r^4}.
\end{eqnarray}
The potential $V$ is complex and the complex frequency $\omega$ is
not separated from the potential. As many authors \cite{Leaver}
\cite{Kokkotas} \cite{Cho} \cite{Simone} have found that such
complex potential can also present correct quasinormal frequencies
provided we use proper boundary conditions.

\section{Numerical Approach}

The QNMs of AdS black holes are usually defined as the solutions
of the relevant wave equations characterized by purely ingoing
waves at the black hole event horizon and vanishing of the
perturbation at the radial infinity. Therefore, the boundary
conditions on wave function $\Psi_s$ (or ${\mathbb{R}}_{s}$) at
the event horizon $(r=r_+)$ and the radial infinity $(r\rightarrow
\infty)$ can be mathematically expressed as
\begin{eqnarray} \label{Bon}
\Psi_s \sim \Delta^{s/2} {\mathbb{R}}_{s} \sim \left\{
\begin{array}{ll} \Delta^{-s/2}e^{-i\omega r_*} &
r\rightarrow r_+, \\
     0 &      r\rightarrow \infty.
\end{array} \right.
\end{eqnarray}
Equations (\ref{wave}), (\ref{Poten}) and (\ref{Bon}) determine an
eigenvalue problem for the quasinormal frequency $\omega$ of the
Dirac field perturbations.

We will calculate the quasinormal frequencies for outgoing Dirac
field (i.e., for the case $s=-1/2$ \cite{Leaver}) by using the
Horowitz-Hubeny approach \cite{Horowitz}. Writing $\Phi_s$ for a
generic wave function as
\begin{eqnarray}
\Phi_s=\Psi_s e^{i\omega r_*},
\end{eqnarray}
then, Eq. (\ref{wave}) can be rewritten as
\begin{eqnarray}\label{wave1}
f^2 \frac{d^2 \Phi_s}{d r^2}+\left(f\frac{d f}{d r}-2i\omega
f\right)\frac{d \Phi_s}{d r}-V\Phi_s=0.
\end{eqnarray}
To map the entire region of interest, $r_+<r<+\infty$, into a
finite parameter range, we change variable to $x=1/r$. Define a
new function $B(x)$ as
\begin{eqnarray}
B(x)=x^2-\frac{1+x_+^2+Q^2x_+^4}{x_+^3}x^3+Q^2 x^4+1,
\end{eqnarray}
then $f$ can be rescaled as
\begin{eqnarray}
f=x^2\Delta=\frac{B(x)}{x^2}.
\end{eqnarray}
In terms of the new variable $x$, Eq. (\ref{wave1}) can be
expressed as
\begin{eqnarray} \label{wave2}
S(x)\frac{d^2 \Phi_s}{d x^2}+\frac{T(x)}{x-x_+}\frac{d \Phi_s}{d
x}+\frac{U(x)}{(x-x_+)^2}\Phi_s=0,
\end{eqnarray}
where
\begin{eqnarray}
S(x)&=&\frac{B(x)^2}{(x-x_+)^2}=\left(\frac{1+x_+^2}
{x_+^3}x^2+\frac{x}{x_+^2}+\frac{1}{x_+}-Q^2 x^3\right)^2, \nonumber \\
T(x)&=&\frac{B(x)}{x-x_+}\left(\frac{d B(x)}{d x}+2i
\omega\right)\nonumber \\ &=&\left(\frac{1+x_+^2}
{x_+^3}x^2+\frac{x}{x_+^2}+\frac{1}{x_+}-Q^2 x^3\right)
\left(\frac{3(1+x_+^2+Q^2x_+^4)}{x_+^3}x^2-2x-4Q^2x^3-2i\omega\right), \nonumber \\
U(x)&=&-V=\frac{B(x)}{4}\frac{d^2 B(x)}{d
x^2}-\frac{s^2}{4}\left(\frac{d B(x)}{d x}\right)^2+is\omega
\frac{d B(x)}{d x}-\lambda^2B(x).
\end{eqnarray}
We can expand $S(x)$ about the event horizon $x=x_+$ as
$S(x)=\sum_{n=0}^{6}S_n(x-x_+)^n$ since $S(x)$ is a polynomial of
degree 6, and similarly for $T(x)$ and $U(x)$.

In order to evaluate the quasinormal frequencies by using the
Horowitz-Hubeny method, we need to expand the solution to the wave
function $\Phi_s$ around $x_+$,
\begin{eqnarray}\label{exp}
\Phi_s=(x-x_+)^\alpha \sum_{k=0}^{\infty}a_k (x-x_+)^k.
\end{eqnarray}
It is easy to show that the index $\alpha$ has two solutions
$\alpha=-\frac{s}{2}$ and
$\alpha=\frac{s}{2}+\frac{i\omega}{\kappa}$. Because we need only
the ingoing modes near the event horizon, that is to say, $\Phi_s$
must satisfy the boundary condition (\ref{Bon}),  we take
$\alpha=-\frac{s}{2}$. Then $\Phi_s$ becomes
\begin{eqnarray}\label{exp1}
\Phi_s=(x-x_+)^{-s/2} \sum_{k=0}^{\infty}a_k (x-x_+)^k.
\end{eqnarray}
Substituting Eq. (\ref{exp1}) into Eq. (\ref{wave2}) we find the
following recursion relation for  $a_n$
\begin{eqnarray}\label{an}
a_n=-\frac{1}{Z_n}\sum_{k=0}^{n-1}\left[\left(k-\frac{s}{2}\right)
\left(k-\frac{s}{2}-1\right)S_{n-k}+\left(k-\frac{s}{2}\right)
T_{n-k}+U_{n-k}\right]a_k,
\end{eqnarray}
where $ Z_n=4n\kappa [(n-s)\kappa -i \omega].$

The boundary condition (\ref{Bon}) at $r\rightarrow \infty$
becomes
\begin{eqnarray}\label{po}
\sum_{k=0}^{\infty}a_k (-x_+)^{k}=0.
\end{eqnarray}
Now the computation of quasinormal frequencies is reduced to find
roots of the Eq. (\ref{po}).  We truncate the sum (\ref{po}) at
some large $k=N$ and then check that for greater $k$ the roots
converge to some true roots, i.e., quasinormal frequencies. In the
series (\ref{an}) and (\ref{po}) each next term depends on all the
preceding terms through the recursion relations, and the roots of
(\ref{po}) suffer a sharp change for a small change on any of the
input parameters, especially for finding  the higher overtone
modes. In order to avoid the ``noisy" we increase the precision of
all the input data and retain 70-digital precision in all the
intermediate process.

\section{Numerical results}

In this section we represent the numerical results obtained by
using the numerical procedure just outlined in the previous
section. The results will be organized into four subsections: the
fundamental QNMs, dependence of the black hole charge, higher
modes, and dependence of the angular quantum number.

\subsection{Fundamental Quasinormal Modes}

\begin{table}
\caption{\label{table1} The fundamental quasinormal frequencies
($n=0$) corresponding to $\lambda=1$ Dirac perturbation of the
RNAdS black holes ($r_+=5, ~10, ~25, ~50, ~75, ~100$) with the
charge $Q=0, ~0.1Q_{ext}, ~0.3Q_{ext}, ~0.5Q_{ext}$.}
\begin{tabular}{c|c|c|c|c}
\hline \hline $ ~~~~~r_+ ~~~~~$ & ~~~~~~~$\omega(Q=0)$ ~~~~~~~&~~~
$\omega(Q=0.1Q_{ext})$ ~~~&~~~ $\omega(Q=0.3Q_{ext})$~~~ &~~~ $\omega(Q=0.5Q_{ext})$ \\
\hline
100 & 199.769-152.016i&198.415-153.628i&187.351-169.030i&182.694-217.611i\\
75  & 149.834-114.012i&148.817-115.220i&140.519-126.770i&137.024-163.204i\\
50  & 99.8993-76.0043i&99.2226-76.8101i&93.6914-84.5094i&91.3576-108.796i\\
25  & 49.9795-37.9952i&49.6411-38.3968i&46.8775-42.2439i&45.7003-54.3789i\\
10  & 20.0744-15.1751i&19.9398-15.3357i&18.8398-16.8675i&18.3406-21.6987i\\
5   & 10.1834-7.54760i&10.1173-7.62708i&9.57673-8.38092i&9.27945-10.7551i\\
\hline \hline
\end{tabular}
\end{table}

\begin{figure}
\includegraphics[scale=0.62]{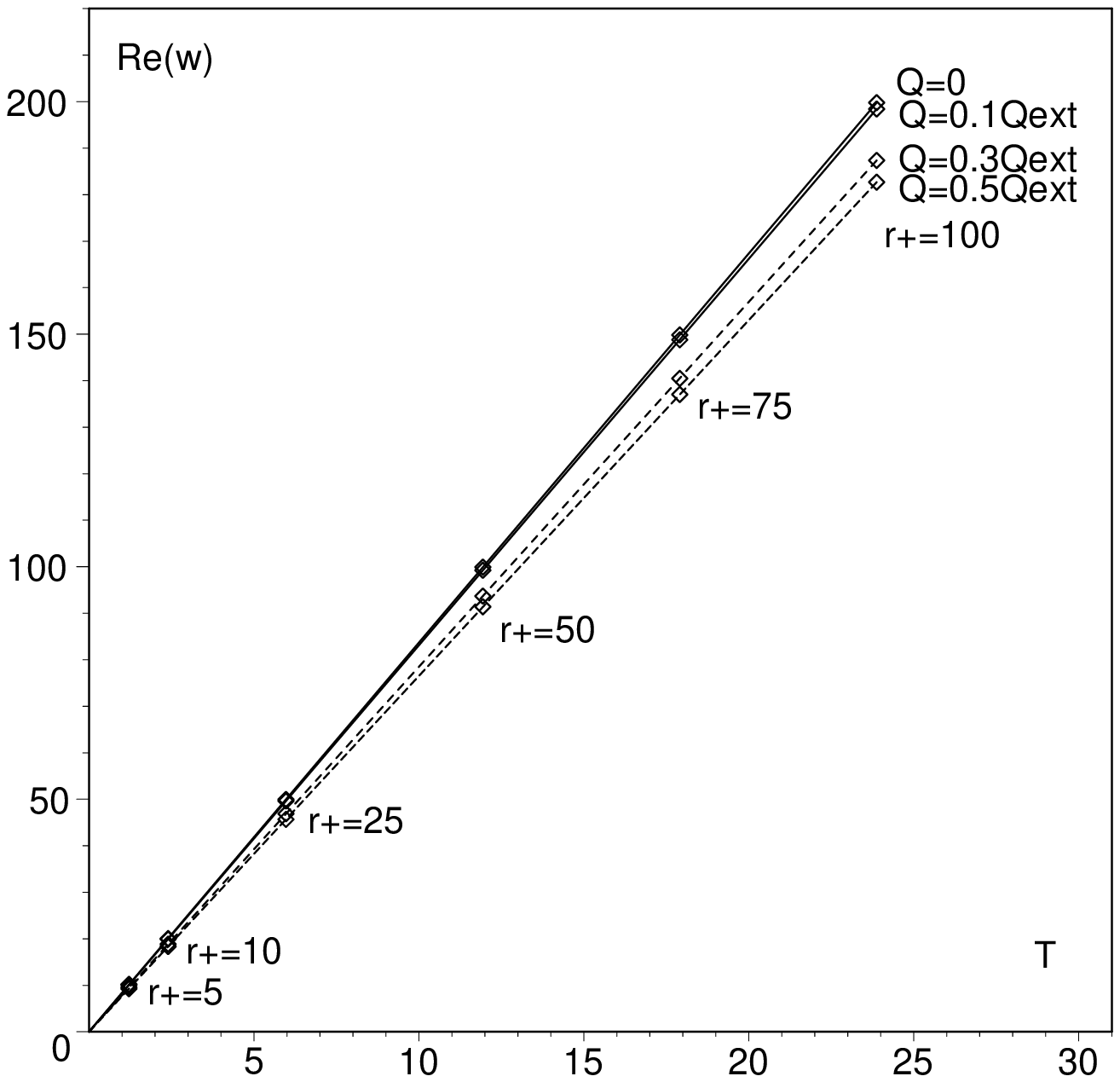}
\includegraphics[scale=0.62]{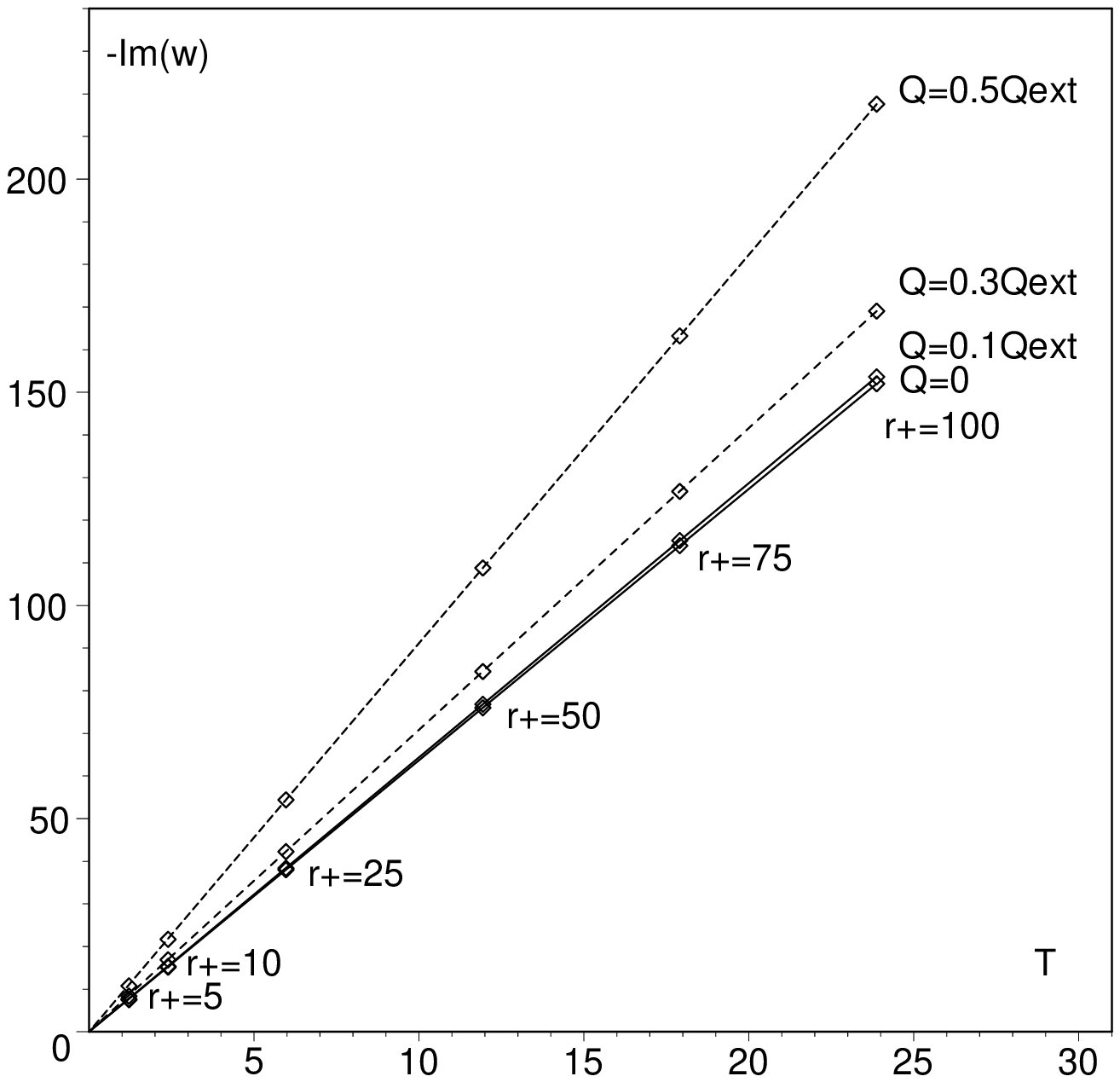}
\caption{\label{fig1} Graphs of fundamental quasinormal
frequencies $\omega$ versus the Hawking temperature $T$  for black
holes ($r_+=5, ~10, ~25, ~50, ~75, ~100$) with $\lambda=1$. The
left figure is drawn for $Re(\omega)$ in which the lines from the
top to the bottom correspond to $Q=0, ~0.1Q_{ext}, ~0.3Q_{ext},
~0.5Q_{ext}$, and the right one for $Im(\omega)$ in which the
lines from the top to the bottom correspond to $Q=0.5Q_{ext},
~0.3Q_{ext}, ~0.1Q_{ext}, ~0$. The figures show that both the real
and the imaginary parts of the frequencies are the linear
functions of $T$.}
\end{figure}

The fundamental quasinormal frequencies ($n=0$) corresponding to
$\lambda=1$ Dirac perturbation for RNAdS black holes ($r_+=5, ~10,
~25, ~50, ~75, ~100$) with the charge $Q=0, ~0.1Q_{ext},
~0.3Q_{ext}, ~0.5Q_{ext}$ are given by table (\ref{table1}) and
corresponding results are drawn in Fig. (\ref{fig1}). From the
table and figures we find that, for the large black holes, both
the real and the imaginary parts of the quasinormal frequencies
are the linear functions of the Hawking temperature, and the lines
are described by
\begin{eqnarray}\label{large}
&& Re(\omega(Q=0))=8.367 T,  \nonumber \\
&& Re(\omega(Q=0.1Q_{ext}))=8.309 T, \nonumber  \\
&& Re(\omega(Q=0.3Q_{ext}))=7.845 T, \nonumber  \\
&& Re(\omega(Q=0.5Q_{ext}))=7.652 T,
\end{eqnarray}
and
\begin{eqnarray}\label{large1}
&&-Im(\omega(Q=0))=6.371 T. \nonumber   \\
&&-Im(\omega(Q=0.1Q_{ext}))=6.439 T,  \nonumber   \\
&&-Im(\omega(Q=0.3Q_{ext}))=7.084 T,  \nonumber   \\
&&-Im(\omega(Q=0.5Q_{ext}))=9.121 T,
\end{eqnarray}
which show that the slope of the lines for the real parts
decreases as the charge increases, but the slope of the lines for
the magnitude of the imaginary parts increases as the charge
increases.

According to the AdS/CFT correspondence, the decay of the Dirac
perturbation can be translated into a time scale for the approach
to the thermal equilibrium in CFT. The time scale is simply given
by the imaginary part of the lowest quasinormal frequency,
$\tau=1/|Im(\omega)|$. From Eqs. (\ref{large1}) we know that for
three-dimensional CFT the time scales are
\begin{eqnarray}\label{timescale}
&&\tau(Q=0)=0.1570 /T. \nonumber   \\
&&\tau(Q=0.1Q_{ext})=0.1553 /T,  \nonumber   \\
&&\tau(Q=0.3Q_{ext})=0.1412 /T,  \nonumber   \\
&&\tau(Q=0.5Q_{ext})=0.1096 /T,
\end{eqnarray}
which show that different black hole charge presents different
time scale.

\subsection{Dependence on the black hole charge }

\begin{figure}
\includegraphics[scale=0.9]{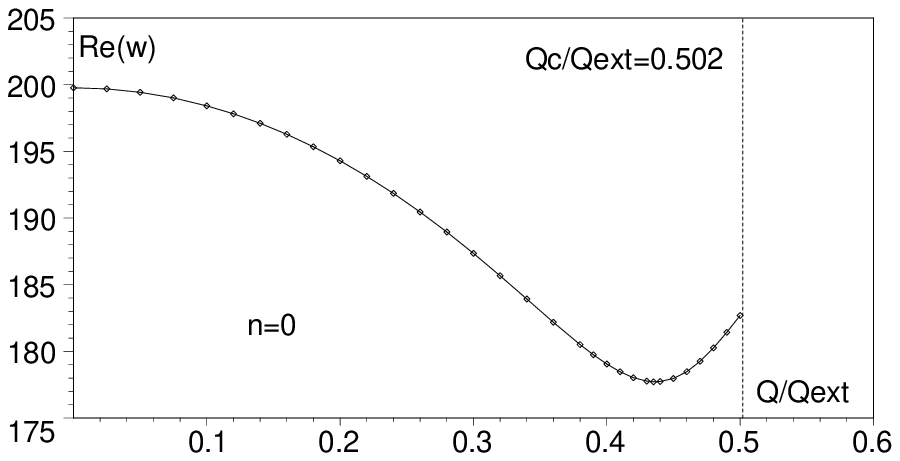}\hspace{0.3cm}%
\includegraphics[scale=0.9]{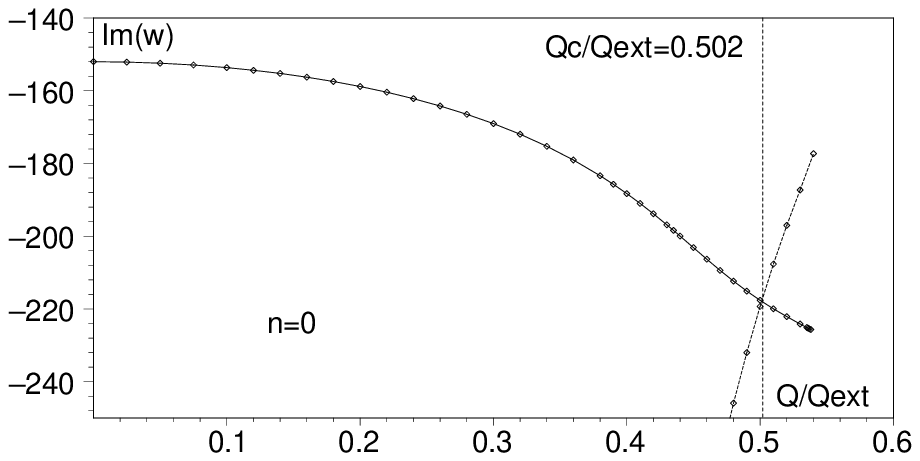} \\ \vspace{0.15cm}
\includegraphics[scale=0.9]{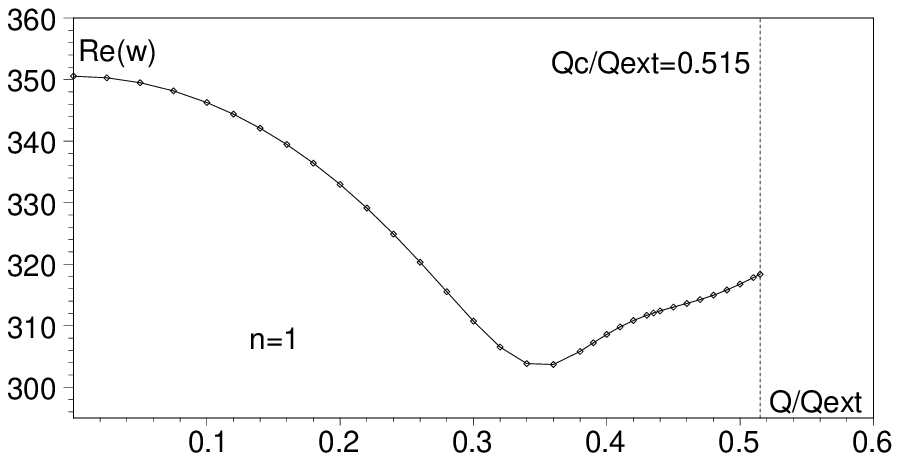}\hspace{0.3cm}%
\includegraphics[scale=0.9]{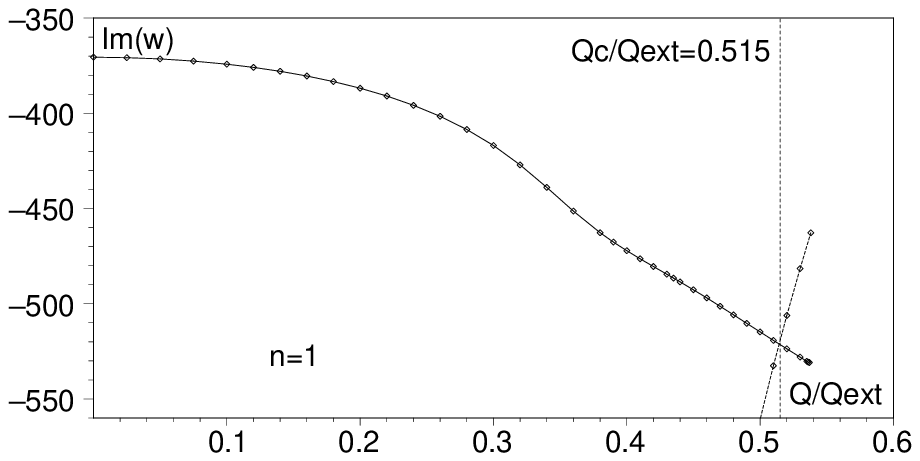} \\ \vspace{0.15cm}
\includegraphics[scale=0.9]{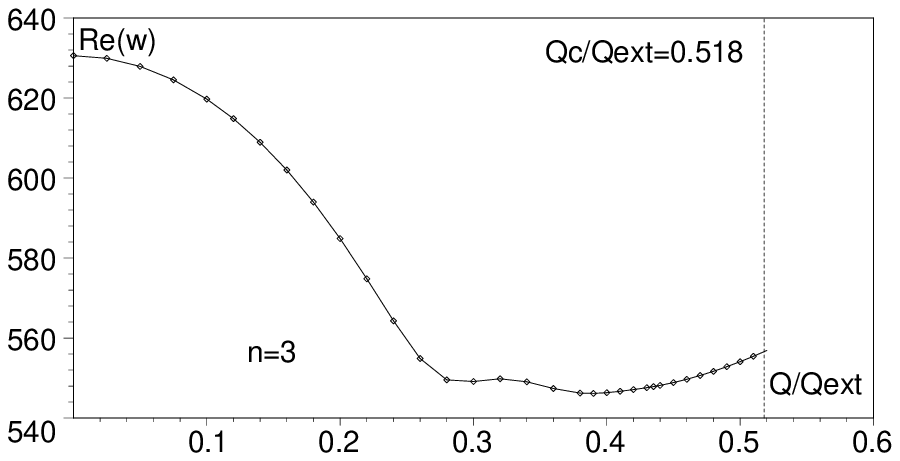}\hspace{0.3cm}%
\includegraphics[scale=0.9]{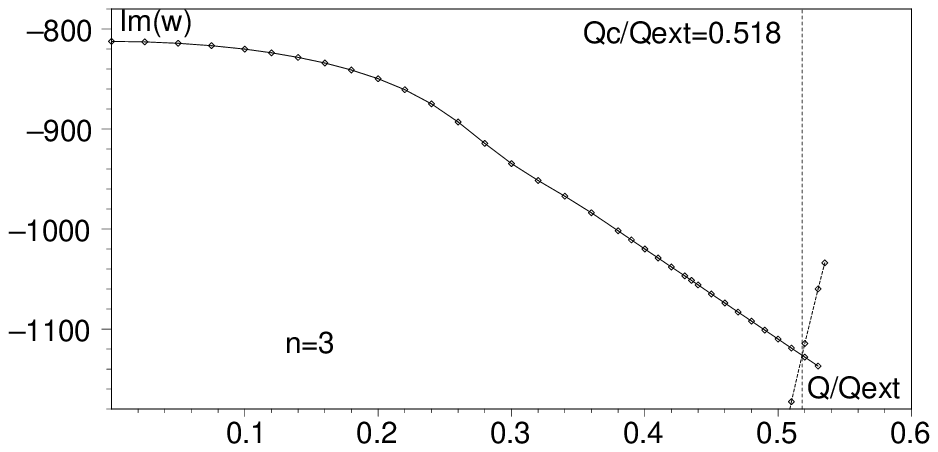}\\ \vspace{0.15cm}
\includegraphics[scale=0.9]{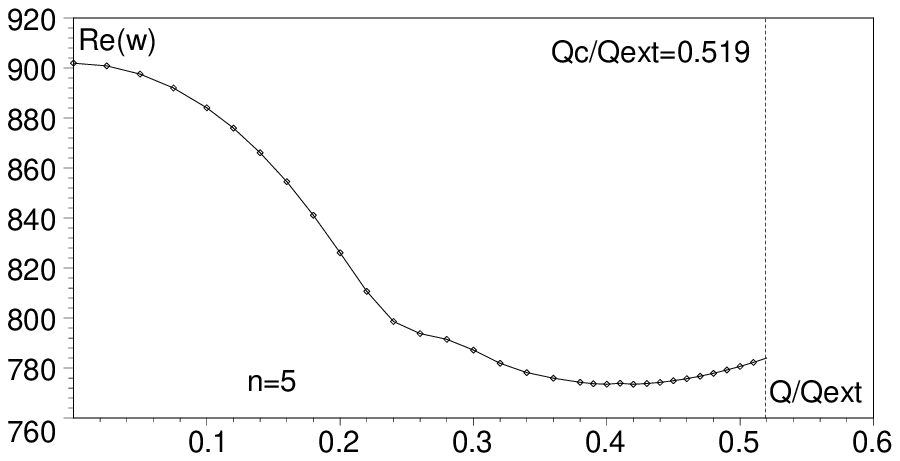}\hspace{0.3cm}%
\includegraphics[scale=0.9]{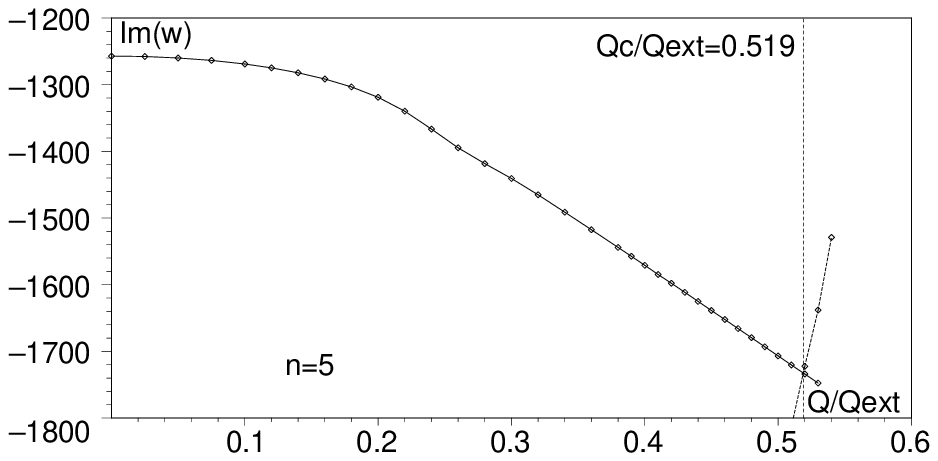}
\caption{\label{fig2}Graphs of $\omega$ with $Q/Q_{ext}$ for $n=0,
1, 3, 5$, $r_+=100$ and $\lambda=1$. The left four figures are
drawn for $Re(\omega)$ which show the ``wiggle behavior". The
right figures for $Im(\omega)$ in which the continuous lines
describe the oscillatory modes and the dashed lines indicate the
non-oscillatory modes. Values of $Q_c/Q_{ext}$  in the graphs
indicate when non-oscillating dominates.}
\end{figure}

The relation between the quasinormal frequencies and the charges
of the black hole for  $n=0, 1, 3, 5$, $r_+=100$ and $\lambda=1$
is shown by Fig. (\ref{fig2}).  We know from figures for
$Re(\omega)$ with $Q/Q_{ext}$ in Fig. (\ref{fig2}) that the local
minima of the ``wiggle" appear as the scalar and gravitational
perturbations \cite{WangB}\cite{Berti2003}. The ``wiggle" becomes
shallower as the overtone number $n$ increases and disappears for
$n\ge 5$. However, we should point out that the curves of
$Re(\omega)$ versus $Q/Q_{ext}$ for the Dirac field are different
from those of the scalar field and gravitational perturbations
\cite{WangB}\cite{Berti2003}.

The figures for  $Im(\omega)$ versus $Q/Q_{ext}$ in Fig.
(\ref{fig2}) tell us that there are two stable classes of
solutions for QNMs when the value of the charge nears $Q_c$, i.
e., the oscillatory modes (the frequencies have both the real and
imaginary parts) and the non-oscillatory modes (the frequencies
are purely imaginary). For the RNAdS black holes with the charge
$Q<Q_c$, the Dirac perturbation decay is dominated by oscillatory
modes. But for the black holes with the charge $Q>Q_c$, the decay
is dominated by non-oscillatory modes. The figures also show us
that the value of the critical value of the charge $Q_c$ slightly
increases as the overtone number $n$ increases.

\subsection{Higher modes}

\begin{figure}
\includegraphics[scale=0.90]{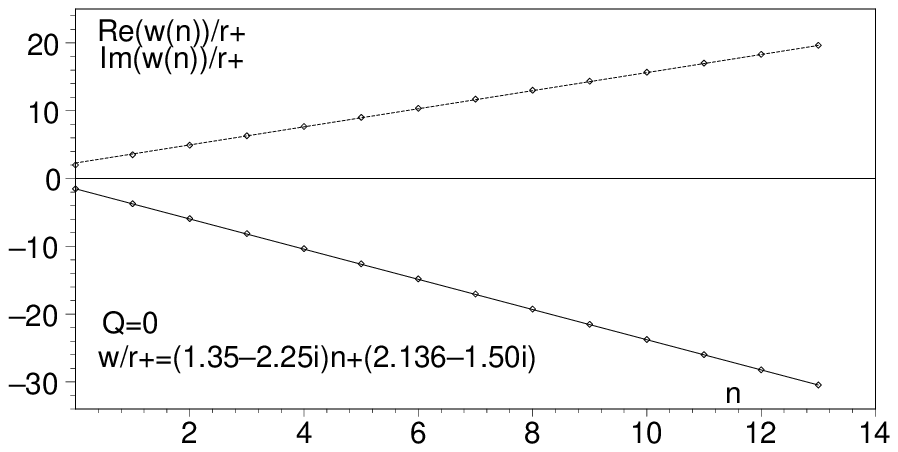}\hspace{0.5cm}%
\includegraphics[scale=0.90]{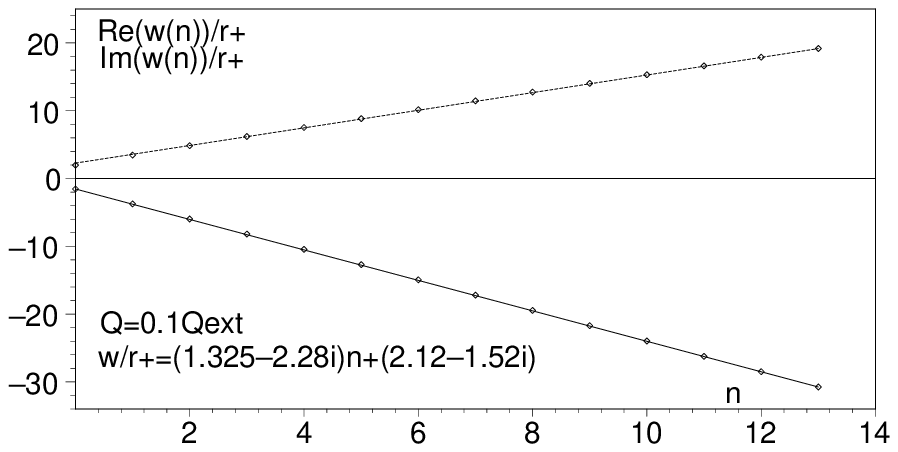} \\ \vspace{0.2cm}
\includegraphics[scale=0.90]{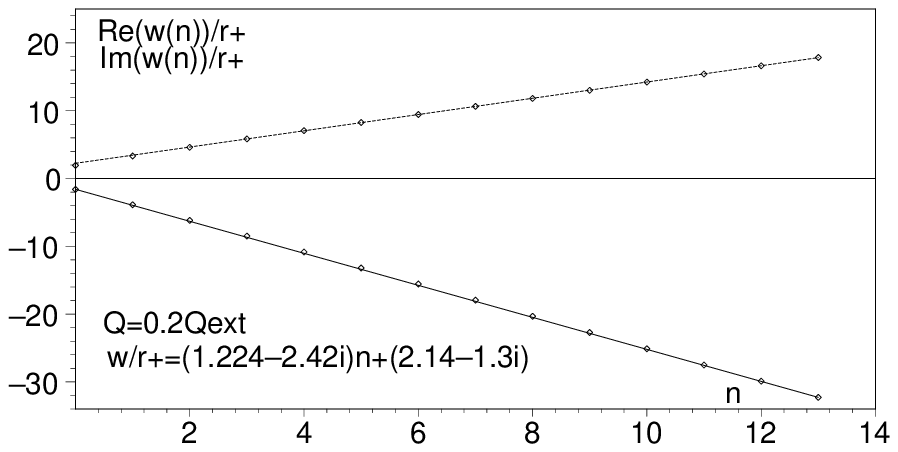}\hspace{0.5cm}%
\includegraphics[scale=0.90]{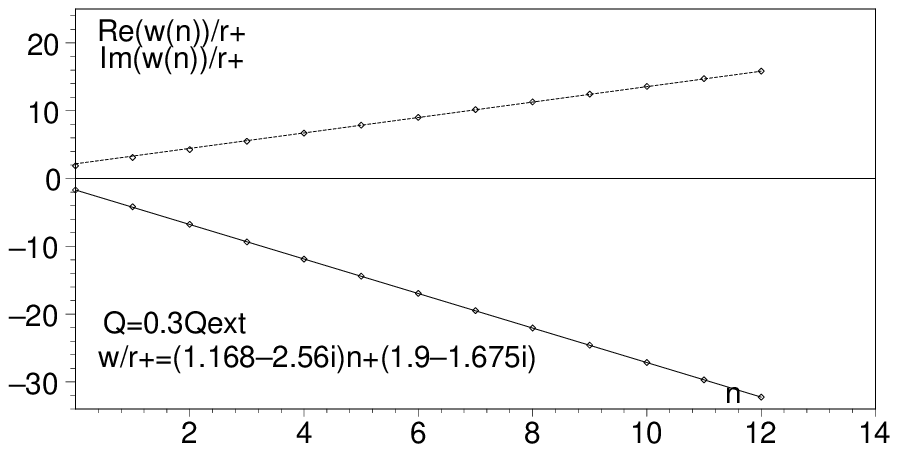}\\  \vspace{0.2cm}
\includegraphics[scale=0.90]{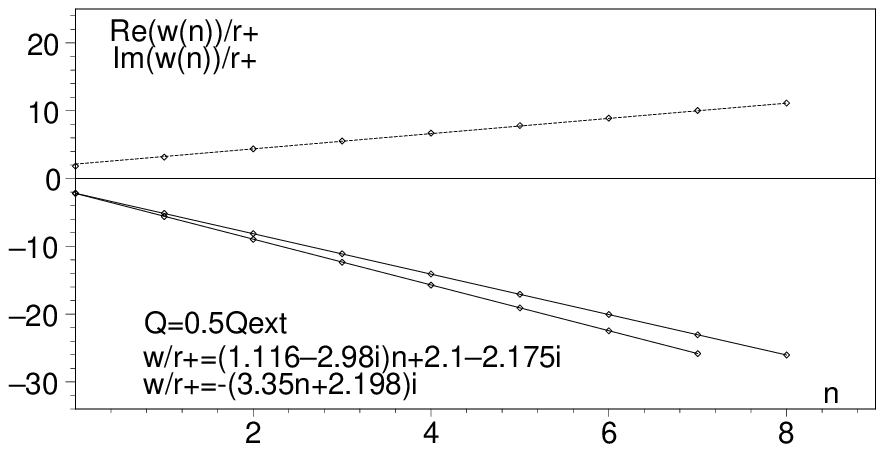}\hspace{0.5cm}%
\includegraphics[scale=0.90]{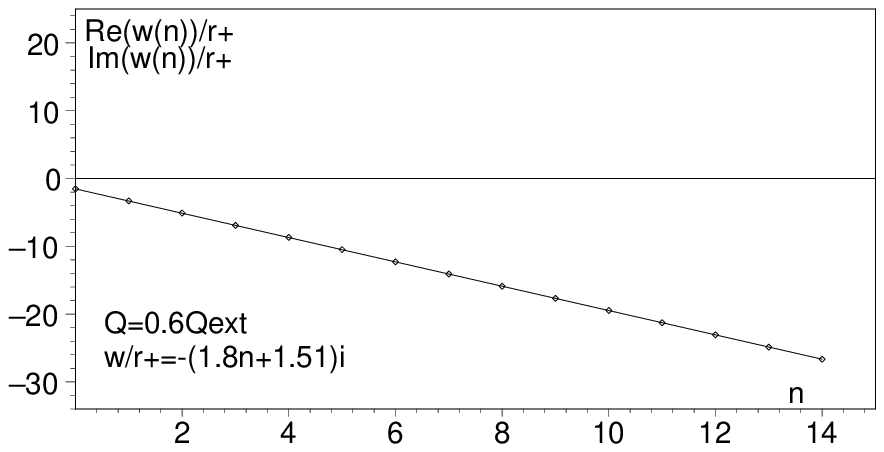}
\caption{\label{fig3}Graphs of $\frac{\omega(n)}{r_+}$ versus $n$
corresponding to $\lambda=1$ Dirac perturbations of a large RNAdS
black hole ($r_+=100$) for $Q=0,~0.1 Q_{ext},~0.2 Q_{ext},~0.3
Q_{ext},~0.5 Q_{ext},~0.6 Q_{ext}$. The dashed lines describe
$\frac{Re(\omega(n))}{r_+}$ and the solid lines are
$\frac{Im(\omega(n))}{r_+}$. It is interesting to note that the
quasinormal frequencies become evenly spaced for high overtone
number $n$. }
\end{figure}

In Fig. (\ref{fig3}) we graph $\frac{\omega(n)}{r_+}$ versus $n$
corresponding to $\lambda=1$ Dirac perturbations of a large RNAdS
black hole ($r_+=100$) for $Q=0,~0.1 Q_{ext},~0.2 Q_{ext},~0.3
Q_{ext},~0.5 Q_{ext},~0.6 Q_{ext}$. The behavior of the QNMs can
be written as $\frac{\omega(n)}{r_+}\sim (1.35-2.25
i)n+(2.136-1.50i)$ for Q=0, $ \frac{\omega(n)}{r_+}\sim
(1.325-2.28 i)n+(2.12-1.52i)$ for $Q=0.1Q_{ext}$,
$\frac{\omega(n)}{r_+}\sim (1.224-2.42 i)n+(2.14-1.30i)$ for
$Q=0.2Q_{ext}$, $\frac{\omega(n)}{r_+}\sim (1.168-2.56
i)n+(1.90-1.675i)$ for $ Q=0.3Q_{ext}$, $\frac{\omega(n)}{r_+}\sim
(1.116-2.98 i)n+(2.1-2.175i)$ and $ -(3.35 n +2.198)i$  for
$Q=0.5Q_{ext}$, $ \frac{\omega(n)}{r_+}\sim-(1.8 n +1.51)i$ for $
Q=0.6Q_{ext}$.

Cardoso et al \cite{Cardoso} found that the scalar quasinormal
frequencies for the large Schwarzschild-AdS black hole become
evenly spaced for high overtone number $n$. Our results extend
their argument to the Dirac quasinormal frequencies of the RNAdS
black hole for fixed value of the charge. The spacings between
frequencies are given by
\begin{eqnarray}
\frac{\omega(n+1)-\omega(n)}{r_+}&\sim& (1.35-2.25 i)
~~~~~~{\textrm{ for}}~~~~
Q=0, \nonumber \\
\frac{\omega(n+1)-\omega(n)}{r_+}&\sim& (1.325-2.28
i) ~~~~{\textrm{ for}}~~~~ Q=0.1Q_{ext},  \nonumber \\
\frac{\omega(n+1)-\omega(n)}{r_+}&\sim& (1.224-2.42 i)
~~~~{\textrm{ for}}~~~~
Q=0.2Q_{ext},  \nonumber \\
\frac{\omega(n+1)-\omega(n)}{r_+}&\sim& (1.168-2.56 i)
~~~~{\textrm{ for}}~~~~
Q=0.3Q_{ext},  \nonumber \\
\frac{\omega(n+1)-\omega(n)}{r_+}&\sim& (1.116-2.98 i)
\nonumber \\
&\sim& (~~~~~~~-3.35 i )  ~~~~{\textrm{ for}}~~~~ Q=0.5Q_{ext}.
\end{eqnarray}
From which we find that, for lowly charged RNAdS black hole, the
real part in the spacing expression decreases as the charge
increases, while the magnitude of the imaginary part increases as
the charge increases.

It is obvious that there are two stable classes of solutions for
the QNMs with  $Q=0.5Q_{ext}$. In the first class the frequencies
have both the real and imaginary parts (oscillatory modes), and in
the second class the frequencies are purely imaginary
(non-oscillatory modes). But there is only an oscillatory mode for
the case with the small charge (say $Q=0.1Q_{ext}$), and there is
only a non-oscillatory modes for the case with the large charge
(say $Q=0.6Q_{ext}$).

\subsection{Dependence on the angular quantum number}

\begin{figure}
\includegraphics[scale=0.52]{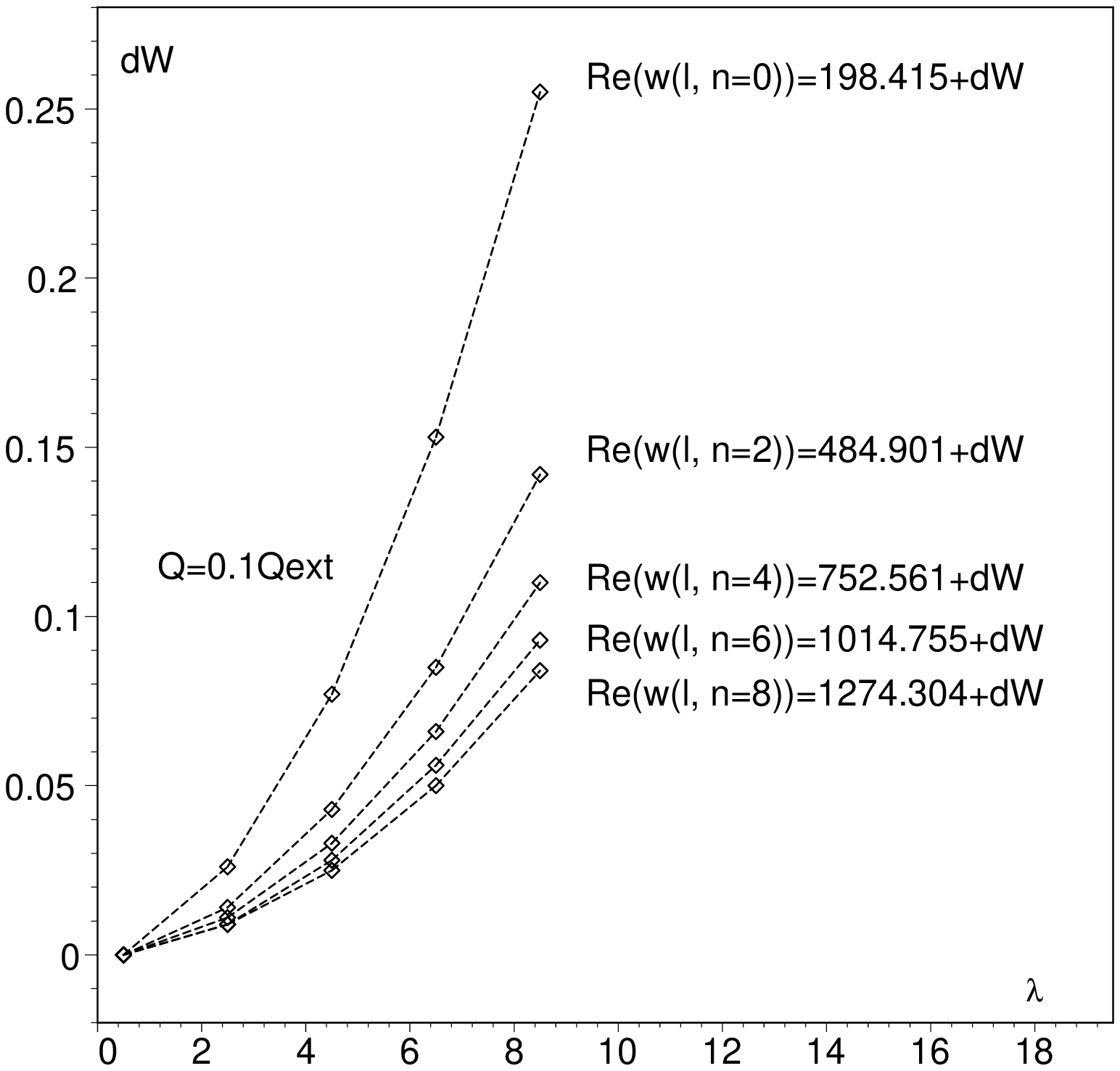}\hspace{0.5cm}%
\includegraphics[scale=0.52]{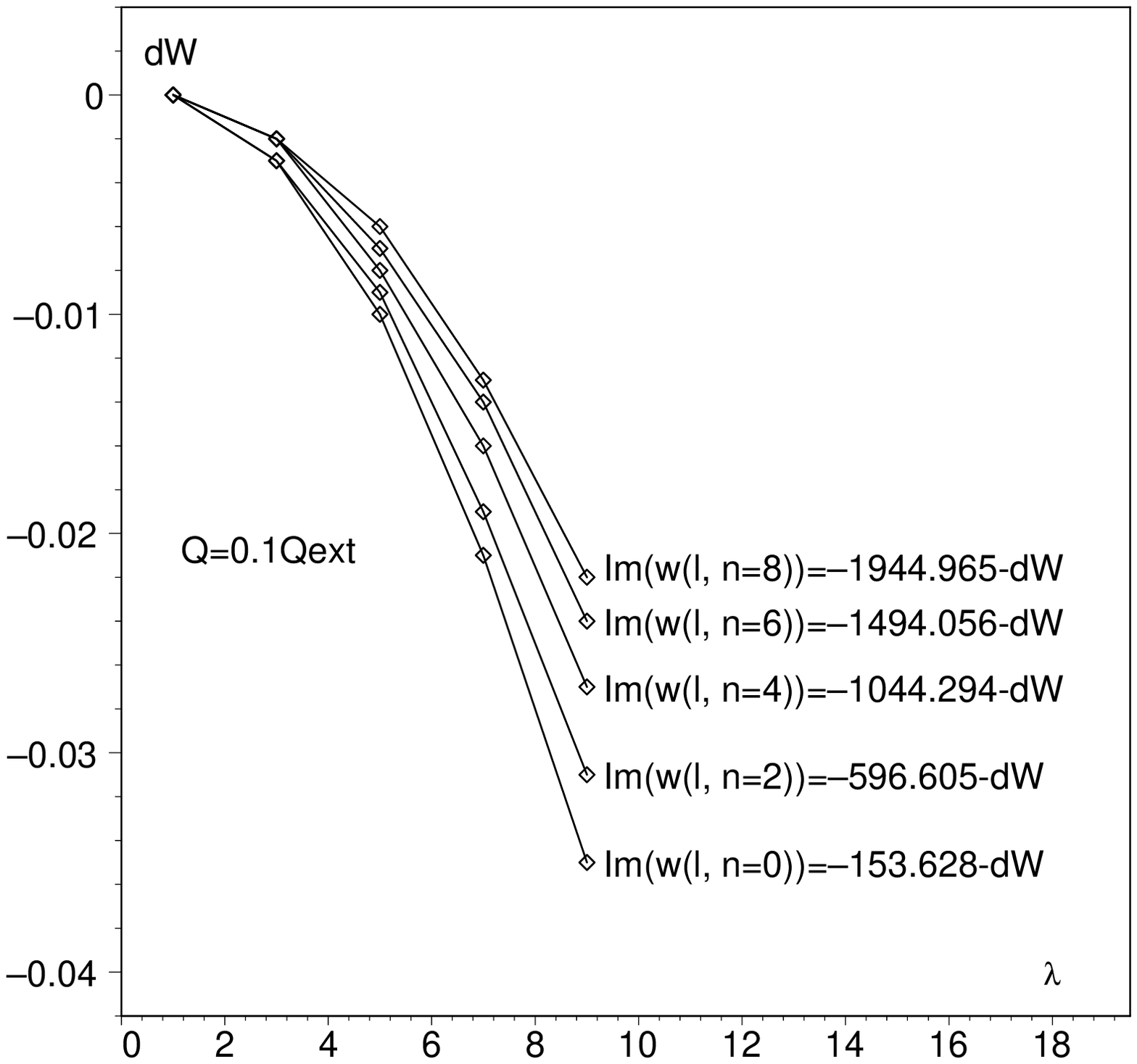} \\ \vspace{0.2cm}
\includegraphics[scale=0.52]{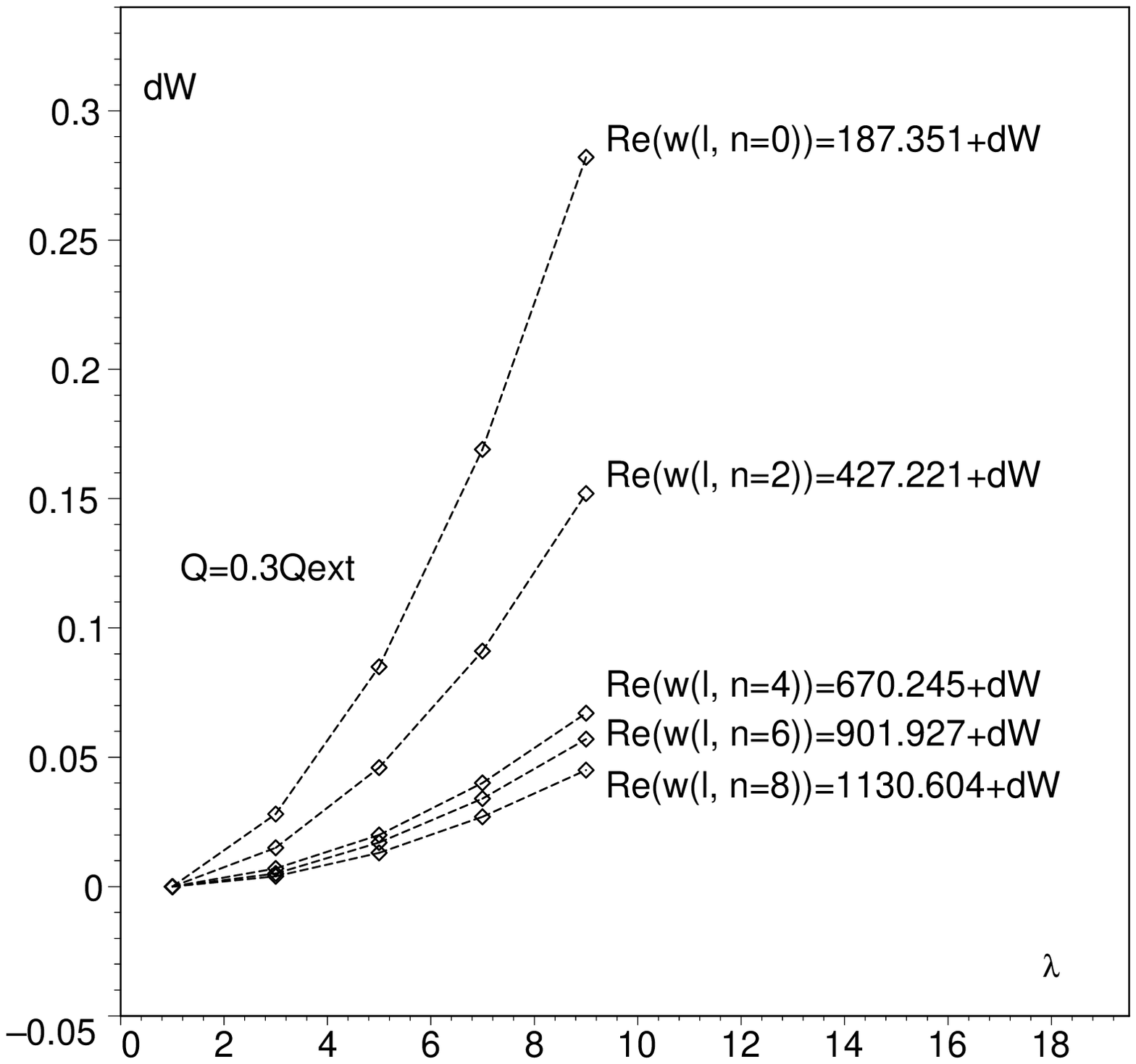}\hspace{0.5cm}%
\includegraphics[scale=0.52]{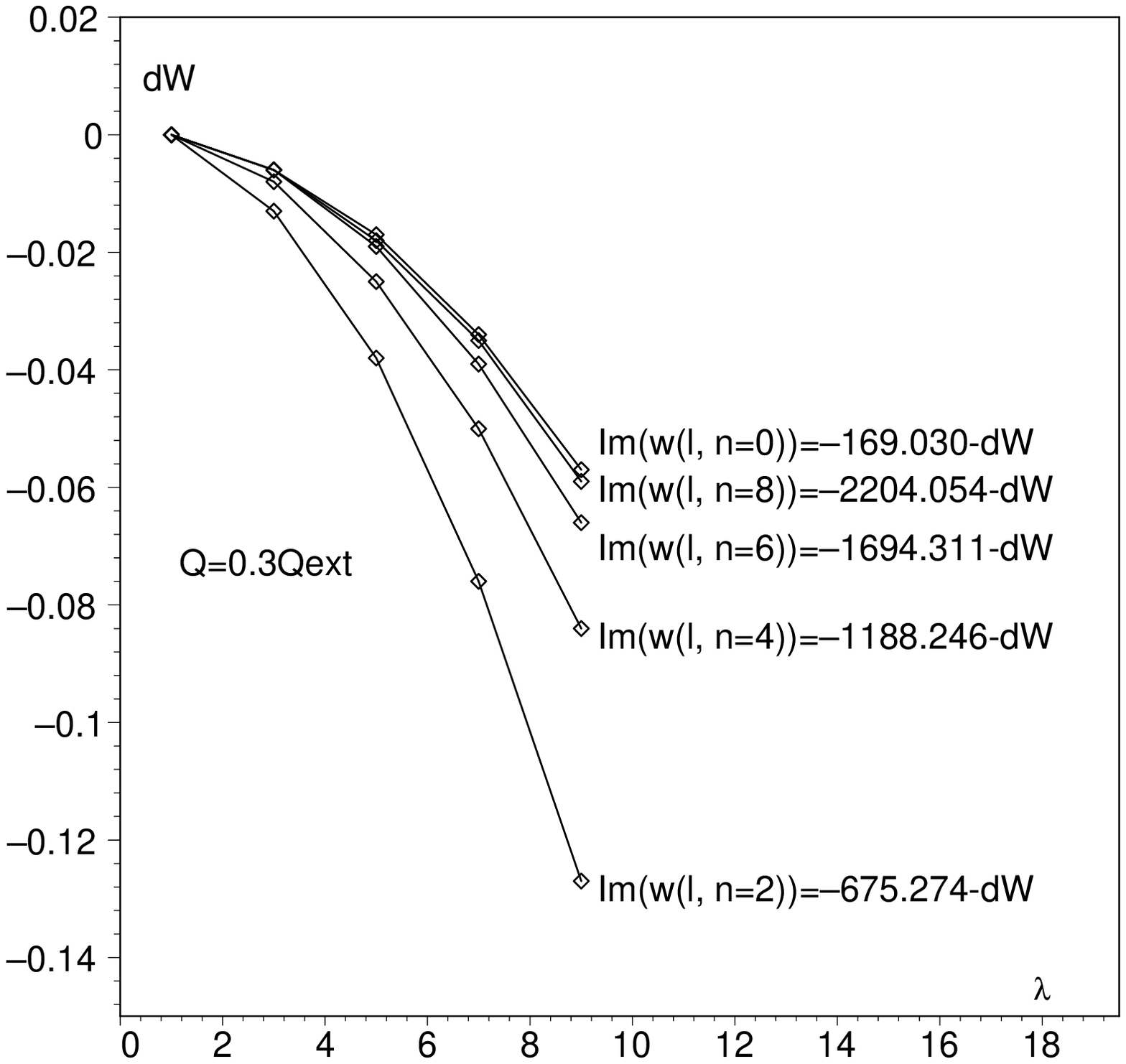}
\caption{\label{fig4} Graphs of the increment $dW$ with $\lambda$
(or $l$) for several value of $n=0, 2, 4, 6, 8$ and $Q=0.1Q_{ext},
~0.3Q_{ext}$. The left two figures are drawn for $dW=Re(\omega(l,
n))-Re(\omega(0, n))$ and the right two are for $dW=-Im(\omega(l,
n))+Im(\omega(0, n))$. The figures show that $Re(\omega)$
increases and $|Im(\omega)|$ decreases with increase of $l$ for
fixed $n$, though the dependence is very weak. Besides, the
figures tell us that the larger the value of the charge $Q$, the
larger the increment $|dW|$ for the imaginary parts.  We also find
that the larger the overtone number $n$, the smaller the increment
$|dW|$ except for the case of $Im(\omega(l,n=0))$ with
$Q=0.3Q_{ext}$. }
\end{figure}

We also study the relation between the quasinormal frequencies and
the angular quantum number $l$ and the results are presented in
figure (\ref{fig4}). We learn from the figures that the
$Re(\omega)$ increases (decreases the oscillatory time scale) and
$|Im(\omega)|$ decreases (increases the damping time scale) as the
angular quantum number $l$ increases for fixed $n$, though
quasinormal frequencies depend very weakly on $l$. Besides, the
figures tell us that for the imaginary parts $|dW|$ increases as
the charge $Q$ increases. We also find that the larger the
overtone number $n$, the smaller the increment $|dW|$ except for
the case of $Im(\omega(l,n=0))$ with $Q=0.3Q_{ext}$.

\section{summary}

The Dirac fields in the RNAdS black hole spacetime are separated
by means of the Newman-Penrose formulism. Then, the quasinormal
frequencies corresponding to the Dirac field perturbations are
studied by using the Horowitz-Hubeny approach and the results are
presented by table and figures. We have shown that: (i) Both the
real and the imaginary parts of the fundamental quasinormal
frequency for the RNAdS black holes ($r_+=5, ~10, ~25, ~50, ~75,
~100$) with the charge $Q=0, ~0.1Q_{ext}, ~0.3Q_{ext},
~0.5Q_{ext}$ are the linear functions of the Hawking temperature.
The slope of the lines for the real parts decreases as the charge
increases, but the slope for the magnitude of the imaginary parts
increases as the charge increases. According to the AdS/CFT
correspondence, the decay of the Dirac perturbation can be
translated into a time scale for the approach to the thermal
equilibrium in CFT. The time scale is simply given by the
imaginary part of the lowest quasinormal frequency,
$\tau=1/|Im(\omega)|$. Thus, for three-dimensional CFT the time
scales are $\tau=0.1570 /T, ~0.1553 /T, ~0.1412 /T, ~0.1096 /T $
for $ Q=0, ~0.1Q_{ext}, ~0.3Q_{ext}, ~0.5Q_{ext}$, respectively,
which show that different charge presents different time scale.
(ii) There are two stable classes of QNMs when the value of the
charge nears $Q_c$, i. e., the oscillatory modes and the
non-oscillatory modes. The Dirac perturbation decay is dominated
by the oscillatory modes for the black hole with the charge
$Q<Q_c$, but the decay is dominated by the non-oscillatory modes
for the black hole with the charge $Q>Q_c$. The value of the
critical value of the charge $Q_c$ slightly increases as the
overtone number $n$ increases. (iii) The Dirac quasinormal
frequencies of the large RNAdS black holes become evenly spaced
for high overtone number $n$. For lowly charged RNAdS black hole,
the real part in the spacing expression decreases as the value of
the charge increases, while the magnitude of the imaginary part
increases as the charge increases. (iv)  The $Re(\omega)$
increases (decreases the oscillatory time scale) and
$|Im(\omega)|$ decreases (increases the damping time scale) as the
angular quantum number $l$ increases for fixed $n$, though
quasinormal frequencies depend very weakly on $l$.

\begin{acknowledgments}This work was supported by the
National Natural Science Foundation of China under Grant No.
10275024 and under Grant No. 10473004; the FANEDD under Grant No.
200317; and the SRFDP under Grant No. 20040542003.
\end{acknowledgments}

\end{document}